\documentclass[preprint]{aastex}
\slugcomment{submitted to {\it Publ. of the Astronomical Soc. of the Pacific}}
\shorttitle{NGC 752}
\shortauthors{AT\&T}

\begin{document}
\title{Extended Str\"omgren Photoelectric Photometry in NGC 752}
\author{Barbara J. Anthony-Twarog and Bruce A. Twarog}
\affil{Department of Physics and Astronomy, University of Kansas, Lawrence, KS 66045-7582}
\email{Electronic mail: bjat@ku.edu,btwarog@ku.edu}

\begin{abstract}
Photoelectric photometry on the extended Str\"omgren system ($uvbyCa$) is presented for 7 giants and 21 main sequence stars in the old open cluster, NGC 752.
Analysis of the $hk$ data for the turnoff stars yields a new determination of the cluster mean metallicity. From 10 single-star members, [Fe/H] =
$-0.06\pm0.03$, where the error quoted is the standard error of the mean and the Hyades abundance is set at [Fe/H] = +0.12. This result is unchanged if all 20 stars within the
limits of the $hk$ metallicity calibration are included. The derived [Fe/H] is in excellent agreement with past estimates using properly-zeroed $m_1$ data, transformed moderate-dispersion spectroscopy, and recent
high dispersion spectroscopy.
\end{abstract}

\section{INTRODUCTION}
Our increasingly sophisticated understanding of the Galaxy's chemical history has been advanced by observational techniques of varying sensitivity and reach.  High-dispersion spectroscopic studies provide the most detailed picture of elemental abundances but are often restricted to sample sizes of dozens (e.g., \citet{spit,luck}) or, occasionally, hundreds of stars \citep{edv93,val5}. Broad-band techniques can simultaneously reach thousands of stars but can be subject to lower parametric sensitivity and more severe coupling of the variables of interest. Somewhere between the exquisite spectral resolution of high-dispersion studies and the more robust broad-band photometric survey tools, the contributions of intermediate-band and narrow-band photometry can provide surprisingly precise estimates of chemical composition and foreground reddening for individual stars, leading to improved determinations of distance and age for individual stars and stellar aggregates in the Milky Way.

Since 1991, we have augmented the highly successful standard Str\"omgren $uvby$H$\beta$ bandpasses with a relatively narrow filter, $Ca$, centered on the near-ultraviolet ionized lines of Calcium \citep{att91}.  Constructed as a color difference analogous to the Str\"omgren $m_1$ index with the $v$ filter replaced by $Ca$, $hk$ retains sensitivity to metallicity for very low metal abundance stars, the original motivation for adding the filter (e.g., \citet{atmp}), to metallicities as high as solar- and Hyades-abundance F dwarfs \citep{at02}.  The extended Str\"omgren system was established by \citet{att91} with photoelectric observations of $\sim$150 stars, and later augmented by nearly 2000 stars with photoelectric observations \citep{att95}, predominantly in the southern hemisphere. Despite the relatively large size of this photoelectric sample, the availability of $hk$ standards is limited to fairly bright field stars scattered around the sky -- not an optimally efficient situation for modern CCD observations.

In large part, the present work was motivated by the desirability of developing $hk$ photometry in a northern open cluster for future use as a secondary standard field for CCD observations.  Although we did not secure as many observations per star as would be desired for secondary standards, the photoelectric photometry discussed in this paper has already been used to supplement field star standards in the reduction of CCD data obtained at the WIYN 0.9 meter telescope in 2003 and 2004; the resulting analysis of $Cavby$H$\beta$ photometry in NGC 2420 is described in a separate publication \citep{att06}.

NGC 752 presents the mixed blessing of being relatively near: it is dominated by moderately bright stars easily within reach of moderate size telescopes while being relatively low in concentration class. While it extends over nearly a degree in diameter (larger than many CCD fields), its bright uncrowded field lends itself to relatively simple, aperture-based measurement strategies.  A review of the cluster's accepted fundamental parameters, as well as a new derivation of the cluster metallicity from the $hk$ index and an updated discussion of the cluster age, (Sec. 4) will follow a brief discussion of the data acquisition and standardization (Sec. 2), and a comparison with previous photometric studies (Sec. 3). Section 5 summarizes our results.

\section{OBSERVATIONS AND REDUCTIONS}
Photoelectric observations with $uvbyCa$ filters were obtained in November, 1999 using a single-channel photometer equipped with a Hamamatsu GaAs photomultiplier mounted at the 24-inch telescope at Mount Laguna Observatory (MLO) in California.  MLO is situated 45 miles east of San Diego in the Cleveland National Forest and is operated by San Diego State University. Stars were observed sequentially through the filters, then again in reverse order. Integration times were set to achieve a count of at least 10,000 above sky through each filter. Observations to determine extinction corrections and coefficients to transform to the standard system were carried out each photometric night.

Following the construction of instrumental magnitudes and indices and corrections for extinction, data from five nights were mapped and averaged onto a common instrumental system. In addition to field-star standards, stars in several open clusters were targeted: NGC 752, NGC 7789, NGC 188 and Melotte 71, but only the stars in NGC 752 were observed to acceptable precision.  

Stars from \citet{att95} were used to arrive at calibrated $V$, $b-y$ and $hk$ indices. In general, separate $b-y$ calibrations should be determined for main sequence stars cooler than $b-y = 0.4$, but no program stars in this color class were observed so this precaution was not observed.  Separate standards for calibrating $m_1$ and $c_1$ indices for cool giants are also necessary. For consistency with other cluster studies, the catalog of \citet{ols93} was used as the source of the cooler standards.  Because the program included giant stars in NGC 752, observations in NGC 752 by \citet{cb70}(CB) were also used in the standardization.  CB's cluster observations are presumed consistent with the catalog of \citet{cb70cat} which forms the primary foundation for Olsen's G-star calibrations.

In all, 34 standard stars were available to calibrate $V, b-y$ and $hk$ indices.  The standard errors of the mean (s.e.m.) of the zeropoint for each calibration equation are 0.002, 0.001 and 0.004, respectively.  A smaller set of just five standards was used to calibrate the $m_1$ and $c_1$ indices for NGC 752 giants, of which three are from the cluster photometry of \citet{cb70}. The standard errors of the mean for the zeropoints of the calibration equations for $m_1$ and $c_1$ were larger, 0.010 for both indices.  Since three of the observed stars may be compared to CB's 1970 values, a separate evaluation of the accuracy of the red giant $m_1$ and $c_1$ indices is possible.  For the three stars in common, our $m_1$ and $c_1$ indices are smaller by an average 0.012 and 0.010 mag than the CB values, though the average differences are not significantly larger than the combined standard errors of the mean differences, 0.016 and 0.013, respectively.

Two tables are presented for the final calibrated photometry: Table 1 contains new $uvbyCa$ photoelectric indices for seven giant stars in NGC 752 while Table 2 contains $V, b-y$ and $hk$ data only for a sample of main sequence stars in NGC 752.  The numbering scheme for stars in NGC 752 follows the WEBDA adopted identifications.  For all of the stars in the present sample, the WEBDA identifications are identical to numbers from \citet{hein}.  The number of nights observed is included in the table following the columns that indicate the standard error of the mean for the indices resulting from the merger of several nights of data or the individual photon-statistical error for single observations.

\section{COMPARISON WITH PREVIOUS PHOTOMETRIC STUDIES OF NGC 752}
Photometry from the present study may be compared to several prior Str\"omgren and broad-band studies in NGC 752, of which \cite{dan} is one of the most extensive.  \citet{dan} sifted through existing photometric, radial-velocity and proper-motion data to construct a detailed picture of membership and binary incidence in NGC 752. Results from nine intermediate-band and broad-band photometric studies were mapped to a common, accepted standard system of $BV$ photometry.  Their compilation may be used to assess the accuracy of the $V$ magnitudes in the present work.  A comparison of the samples yields an overlap of all 28 stars. The mean difference in the sense (Daniel - MLO) is $-0.010 \pm 0.016$ where the error refers to the standard deviation of a single star.  The $V$ magnitudes for star H 206 and H 235 in the present study are fainter than in the \citet{dan} compilation by more than 0.05 mag.  H 235 was first noted as a variable by \citet{d88} and has been analyzed as an eclipsing binary by \citet{MI95}. H 206 is a spectroscopic binary with rapid rotation \citep{dan}, but has not exhibited variability in past studies. Excluding these two stars, the average residual in $V$ is reduced to $-0.007 \pm 0.011$ mag, with the magnitudes in the current investigation being fainter.

As part of an effort to assess the uniformity of Str\"omgren photometry, \citet{jt} observed stars in NGC 752, seven of which overlap with our sample.  On average, the $V$ magnitudes for the present study are 0.028 mag fainter than those of \citet{jt} with a standard deviation of 0.018.  One star is again conspicuously discrepant, H 206.  Excluding that star, the mean difference in the sense (JT - MLO) is not much changed, $-0.022$, but the dispersion is tighter, 0.005, indicating greater significance for the mean difference.  The $b-y$ colors are generally quite similar in the two samples, with a mean difference in the same sense of $-0.012 \pm 0.009$.

The earlier $uvby$ study by \citet{tw83} affords a larger sample of 11 turnoff and main sequence stars for comparison.  Mean differences for $V$ and $(b-y)$ are $-0.012 \pm 0.020$ and $-0.007 \pm 0.008$, respectively, in the sense of (TW83 - MLO).  Once again, star H 206 appears fainter in the present survey by about 0.065 mag.  If that star is eliminated from the comparison, the mean difference for $V$ drops to $-0.007 \pm 0.009$, the same difference found with the larger composite sample of \citet{dan}.   A very similar result is obtained by comparing the present photometry to that of \citet{cb70}.  For 20 main sequence stars in common, the mean difference in the sense (CB - MLO) for $b-y$ colors is $-0.009 \pm 0.009$.  

\section{BASIC PROPERTIES OF NGC 752}
\subsection{Reddening and Metallicity}
With a distance of less than 500 pc, NGC 752 is unlikely to have substantial foreground reddening but determining the foreground color excess is a necessary step before precise determination of the cluster distance and metal abundance are possible.  The maximum reddening in the direction of the cluster has been estimated by \citet{schl} to be $E(B-V)=0.054$, a plausible upper limit for the cluster reddening. 
With the ability to exclude binaries and probable non-members, \citet{dan} were able to provide accurate estimates for the foreground reddening based on several different photometric approaches.   Their adopted value for $E(B-V)$ of $0.035 \pm 0.005$ was consistent with a weighted photometric estimate for [Fe/H] of $ -0.15 \pm 0.005$ based upon UBV, Str\"omgren, DDO, Geneva and Washington photometry and the metallicity calibrations at that time. For the remainder of this discussion, we have adopted a reddening value for NGC 752 consistent with the \citet{dan} result, or $E(b-y)=0.027$.  

Estimates of NGC 752's chemical abundance, summarized by \citet{dan} and \citet{atat} have ranged from slightly subsolar to nearly solar.  Among the earlier studies, in addition to identifying some chemically peculiar stars, \citet{rfg} was apparently interested in investigating claims of metal paucity for NGC 752 advanced by \citet{bb} and \citet{gw}.  The results of \citet{rfg} indicated an essentially solar abundance, though the sizes of the error bars on many of these published abundances were rather sketchy.  A more recent estimate of approximately solar abundance has been derived by \citet{sest}, based on a high dispersion spectroscopic study of 18 main sequence G stars. The analysis also references observations of the sun and Hyades members of comparable spectral class.  Their derived abundance for NGC 752 of [Fe/H]$=0.01 \pm 0.04$ is essentially solar; for reference, their abundance for the Hyades is $0.14 \pm 0.06$.  

Other high dispersion spectroscopic surveys have indicated a modest metal deficiency, including \citet{hobbs} who derived an abundance of [Fe/H]$=-0.09$, consistent with a reddening value of $E(B-V)=0.04$. Their sample of eight main sequence members was a subset of a larger sample of 19 F stars analyzed by \cite{loucaty}.  
Photometric surveys have tended to indicate a modest subsolar abundance.  \citet{dan} reviewed a variety of photometric methods, including Washington photometry, $UBV$ and Geneva systems, with estimates for [Fe/H] as low as $-0.20$.

More recently, by constructing a homogeneous and self-consistent dataset of cluster abundances, \citet{taat} used merged abundances from moderate-dispersion spectroscopy and DDO photometry to study the detailed structure of the galactic abundance gradient and its evolution. The DDO abundance estimates were based upon a recalibration of the DDO system tied to high dispersion spectroscopic abundances for over 400 field giants \citep{tat96}. Major contributions to the foundational abundance scale were formed by the open cluster sample studied by \citet{fj} and \citet{tho} and summarized in \citet{fr95}. This data set has since been revised and expanded by \citet{fj02}. 

\citet{taat} discussed the relationship between abundance determinations of \citet{fr95} and their adopted abundance scale. The resulting transformation was neither a simple offset nor a straight line, but instead required a systematic increase in [Fe/H] for the more metal-poor clusters, followed by a linear transformation for clusters near solar abundance and higher; the most metal-rich clusters were reduced in [Fe/H]. The spectroscopic data for NGC 752 \citep{fj} produced [Fe/H]$ = -0.16$; transformed to the DDO-defined scale, \citet{taat} found [Fe/H]$ = -0.04$. The direct DDO observations of the cluster giants produced [Fe/H]$ = -0.16$. The final abundance, [Fe/H]$= -0.09$, represents a weighted combination of the DDO and spectroscopic results. 

Since the revision by \citet{fj02}, this discussion of the relationship between the DDO scale and that based upon moderate-dispersion spectroscopy has been updated somewhat by \citet{att06}. The more recent compilation by \citet{fj02} yields a result [Fe/H]$ = -0.18 \pm 0.04$ from 9 stars.  Using the revised transformation of \citet{att06}, the abundance for NGC 752 \citep{fj02} on the cluster metallicity scale of \citet{taat} is [Fe/H]$ = -0.08 \pm 0.04$.

Although we lack new H$\beta$ photometry with which to make an independent assessment of the cluster foreground reddening, with newly acquired $hk$ photometry in the cluster and access to prior $uvby$H$\beta$ photometric studies, we can now add new data to the photometric abundance estimates for NGC 752.
Our observations yielded $hk$ indices for 21 main sequence members in NGC 752. Of these, ten stars will be used for the computation of abundance, following the exclusion of seven spectroscopic binaries noted by \citet{dan} and designated as a note in the final column of Table 2.  Four other stars are excluded for various reasons.  For H 192, no $uvby$H$\beta$ photometry is available from \citet{cb70}. Star H 209, a member but an obvious blue straggler outside the temperature range of the metallicity calibration, was excluded in the analysis of $hk$ indices, as was star H 193 for which spectroscopic and photometric measures indicate an anomalously high metal abundance, consistent with the designation by \citet{rfg} as a chemically peculiar star.  The published $m_1$ index for H 139 indicates either a photometric error or anomaly and it has been excluded from both photometric abundance analyses. 

For the remaining 10 stars, we appeal to the [Fe/H], $\delta hk$ calibration presented in \citet{atmp} and modified in \citet{atat}.  The modification replaces $b-y$ with H$\beta$ as the temperature proxy for a standard $hk$,temperature relation. The relationship between [Fe/H] and $\delta hk$ at the color of the NGC 752 turnoff is [Fe/H]$ = -3.4 \delta hk(\beta) + 0.12$, where the zeropoint forces consistency with a Hyades metallicity of $+0.12$. For these 10 stars, the average $\delta hk = 0.055$ with a standard deviation of 0.025, implying [Fe/H]$ = -0.06 \pm 0.03$ where the error describes the standard error of the mean value. Note that the estimated [Fe/H] is virtually identical if all stars are included in the average, with the exception of the blue straggler, which lies outside the calibration range of the index.

The abundance determination from the $m_1$ indices is straightforward after applying corrections for reddening, again comparing to standard relations and constructing a $\delta m_1$ index related to [Fe/H].  Since new $m_1$ indices were not obtained in the current study, a key question was the choice of the published values to use in the comparison. The data from the original study of \citet{cb70} was adopted because the $m_1$ discrepancy noted by \citet{tw83} almost certainly originates in the zero-point of the later paper \citep{ni88,jt}. 

As in previous cluster studies (e.g., \citet{at36}), the calibration defined by \citet{ni88} was used to derive an abundance from $\delta m_1$ values constructed relative to a standard relation with H$\beta$ as the defining temperature index, resulting in [Fe/H]$ = -0.05 \pm 0.03$ for the selected ten stars. If the $m_1$ data of \citet{cb70} and \citet{tw83} are placed on a common system defined by \citet{cb70}, the mean [Fe/H] from 31 stars is 
$-0.08 \pm 0.02$ (s.e.m.) \citep{dan}. From 26 stars, \citet{ni88}, using the same calibration adopted here, found [Fe/H]$ = -0.05 \pm 0.03$ (s.e.m.).  

We close this section by re-emphasizing the value of using H$\beta$ as the primary temperature index in deriving the metallicity. Since H$\beta$ is independent of reddening and metallicity and not included within the definition of either $hk$ or $m_1$, as $b-y$ is, one does not face the issue of correlated errors in metallicity and temperature.
Moreover, one avoids any questions regarding the effect of a zero-point error in $b-y$. The $b-y$ indices found here are typically redder than those of \citet{cb70} and \citet{tw83}. This offset would be removed in the process of calculating the reddening using H$\beta$, i.e., a zero-point offset in $b-y$ would translate into an offset in $E(b-y)$, leaving the reddening-free index $(b-y)_0$ correct. For the metallicity determination, an error in $E(b-y)$
would have no impact on H$\beta$, but would alter $m_0$ and $hk_0$. Fortunately, the slopes defining the
error propagation for $m_1$ and $c_1$, in the sense $\Delta$[Fe/H]/$\Delta$$E(B-V)$, are only 2.5 and 0.5, respectively. An
error of 0.01 in $E(B-V)$ translates into an error in [Fe/H] of less than 0.03 dex from $m_1$ and 0.005 dex for $hk$.

\subsection{Cluster age and giant branch features}

Discussion of a cluster's age may be usefully revisited whenever estimates of its reddening or metal abundance are altered, or when significant updates in the construction of theoretical isochrones become available.  The dataset compiled by \citet{dan}, comprising photometry as well as membership information based on proper motions, has been utilized in a number of comparisons to theoretical isochrones, including \citet{dan} themselves.  By estimating the effect of NGC 752's slightly subsolar metallicity on the isochrone comparison, \citet{dan} arrived at an age estimate of $1.7 \pm 0.1$ Gyr by comparison to models from \citet{mmm}. The same dataset was used by \citet{ddgp} to compare to isochrones incorporating improved OPAL opacities;  they adopted a slightly lower value for [Fe/H]$ = -0.27$ and arrived an older age estimate of $2 \pm 0.4$ Gyr. Yet another use of the \citet{dan} photometry was incorporated by \citet{at36} in a discussion of the open cluster, NGC 3680, highlighting the very similar morphological characteristics of the two clusters' color-magnitude diagrams.  NGC 3680 and NGC 752 exhibit very similar differences in magnitude and color between their giant branch clumps and unevolved main sequence, consistent with moderately similar ages and chemical composition.  In a comparison to solar abundance isochrones published by \citet{pad} with an estimated offset to accommodate NGC 752's subsolar metallicity, an age of $1.55 \pm 0.1$ Gyr was derived consistent with an adopted reddening of $E(B-V) = 0.03$.  In Figure 1, we add another comparison of the \citet{dan} $BV$ photometric data to theoretical isochrones, in this case models published by \citet{VdBB} for a scaled solar composition [Fe/H]$ = -0.11$.  The isochrones have been shifted by amounts consistent with an apparent distance modulus of 8.3 and reddening $E(B-V) = 0.035$ for the comparison.  An age of 1.6 Gyr appears to be consistent with the comparison.

Direct comparison of NGC 3680 and NGC 752 highlights an interesting feature shared by the two clusters,that of apparently bimodal giant branch clumps, a feature explored at length for NGC 752 by \citet{merm}.  \citet{merm} provide additional membership information and spectroscopic binary identification for red giants in the cluster.  They were able to verify the membership credentials of stars in the bluer clump at the level of the red giant branch and noted that available lithium abundance measurements by \citet{psh} and \citet{gil} indicated weak or absent lithium in the bluer clump giants.  In contrast, the only red giants with detectable lithium are found along what appears to be the first ascent red giant branch.  \citet{merm} also puzzled over the stars bluer and fainter than the blue side of the clump, with no obvious explanation found for their position.

The recently published isochrones of \citet{VdBB} not only provide a fine grid of abundances but transformation to 
Str\"omgren indices.  As an illustrative test, we present a second comparison of their isochrones for a scaled solar abundance [Fe/H]$ = -0.11$ and ages 1.4, 1.6 and 1.8 Gyr with photometry from tables 1 and 2.  As for Figure 1, the isochrones have been shifted by 0.027 and 8.3 to match the observational data reddened by that amount and the apparent distance modulus of the cluster. Photometric data for an additional member giant, H 311, is presented as well, from \citet{cb70}, with this point denoted by an asterisk in Figure 1.  In the figure, open symbols denote stars identified as binaries by \citet{dan} or \citet{merm}.  For the giant stars, triangles denote the positions of stars with detectable Lithium, H 77 and H 208 \citep{psh, gil}. Star H 27 is a member \citep{merm} but no lithium abundance information is available.  While the sample of stars near the turnoff is meager and photometry on the unevolved main sequence is lacking, the comparison to isochrones does indicate concordance
with an age estimate of 1.6 Gyr for the adopted reddening and composition.  

\section{CONCLUSIONS}
Photoelectric photometry on the extended Str\"omgren system has been obtained of stars covering a wide range in temperature and luminosity class in the old open cluster NGC 752. While the primary motivation for the observations
is the establishment of potential standard stars for CCD applications in the northern sky, the new data also supply an opportunity for evaluating the cluster metallicity, a subject of interest for the last 20 years. 

Using metallicity calibrations for the $hk$ index derived and refined over the last decade, the mean [Fe/H] from 10 stars, selected as single stars without evidence for spectroscopic anomalies, becomes --0.06$\pm$0.03 (s.e.m.), on a scale where the Hyades has [Fe/H] = +0.12.
If all 20 stars within the calibration range of the index are used, the mean and the dispersion remain the same. This is slightly higher than the value ([Fe/H] = $-0.15\pm$0.05) found in the comprehensive study of NGC 752 by \citet{dan}, and
the same within the errors as the abundance derived in \citet{taat} from the weighted average of DDO photometry and transformed moderate-dispersion spectroscopy ([Fe/H] = $-0.09 \pm 0.02$).
These values are also in excellent agreement with the previous $uvby$ analyses of over two dozen stars at the cluster turnoff by both \citet{cb70} and \citet{ni88}, whose photometry generates
[Fe/H]$ = -0.08 \pm 0.02$ and [Fe/H]$= -0.05 \pm 0.03$, respectively. The only large, high-dispersion spectroscopic survey of the cluster
is that of \citet{sest}, who finds [Fe/H]$ = -0.01 \pm 0.04$, on a Hyades scale of +0.12. This abundance is clearly consistent within the uncertainties with the estimates
obtained by earlier $uvby$ studies and recent cluster compilations when transformed to a common scale. Thus, the best
estimate for the metallicity of NGC 752 that encompasses the photometric and spectroscopic results is [Fe/H]$ =
-0.05 \pm 0.05$, on a scale where the Hyades has [Fe/H] = +0.12. 

\acknowledgements
In addition to the support of the Department of Physics and Astronomy and the General Research Fund of the University of Kansas, BJAT owes a great deal to the hospitality of the staff and students at Mount Laguna Observatory and the astronomy program at San Diego State University.   Extensive use was made of the SIMBAD database, operating at CDS, Strasbourg, France  and the WEBDA database maintained at the University of Geneva, Switzerland.

\clearpage

\figcaption[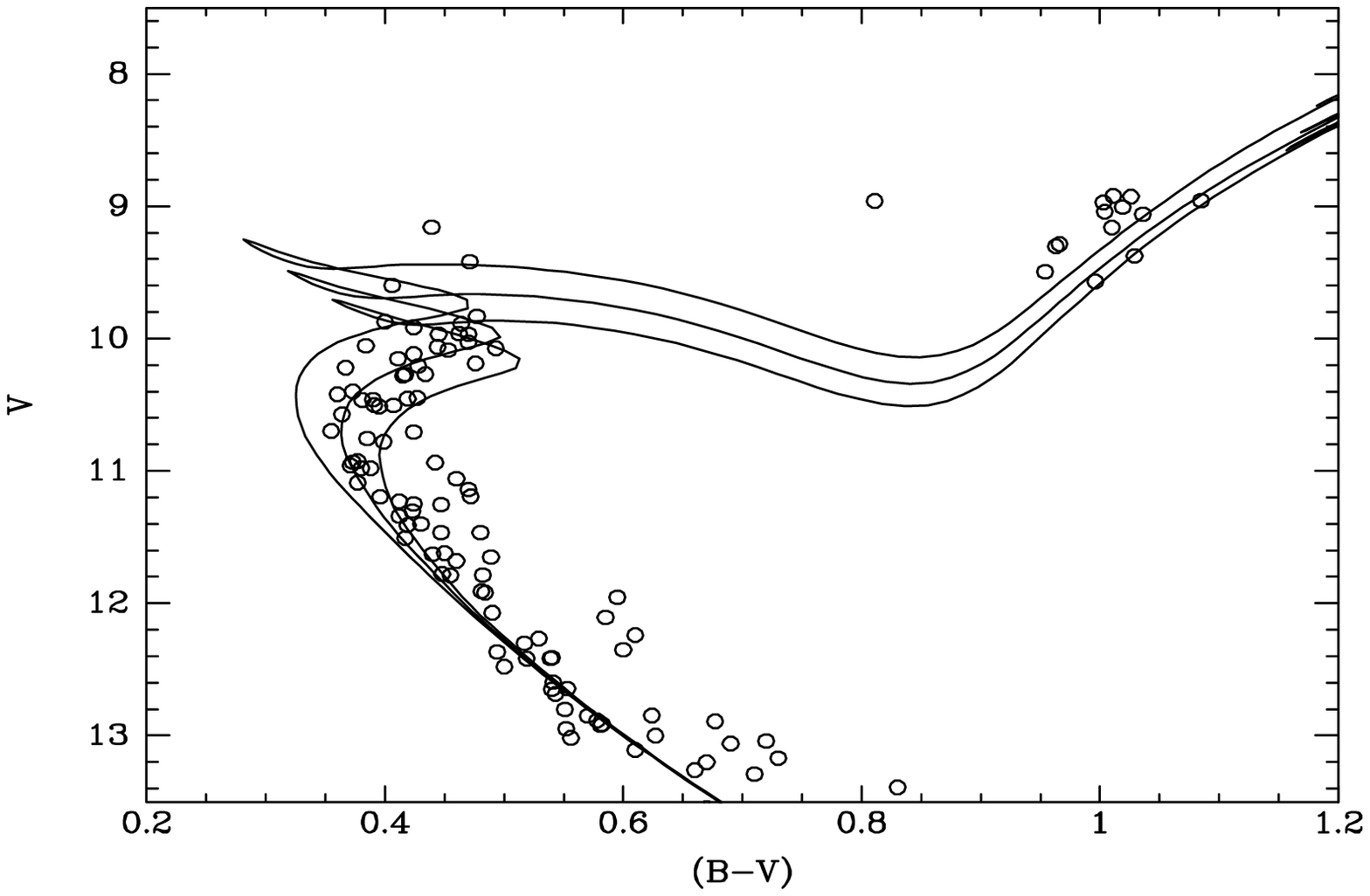]{$BV$ color-magnitude diagram for NGC 752, with data from \citet{dan} and isochrones from \citet{VdBB} for a scaled-solar composition with [Fe/H]$ = -0.11$ and ages 1.4, 1.6 and $1.8 \times 10^9$ years. The isochrones have been adjusted by 8.3 magnitudes and 0.035 magnitudes in color to match the data.\label{fg1}}

\figcaption[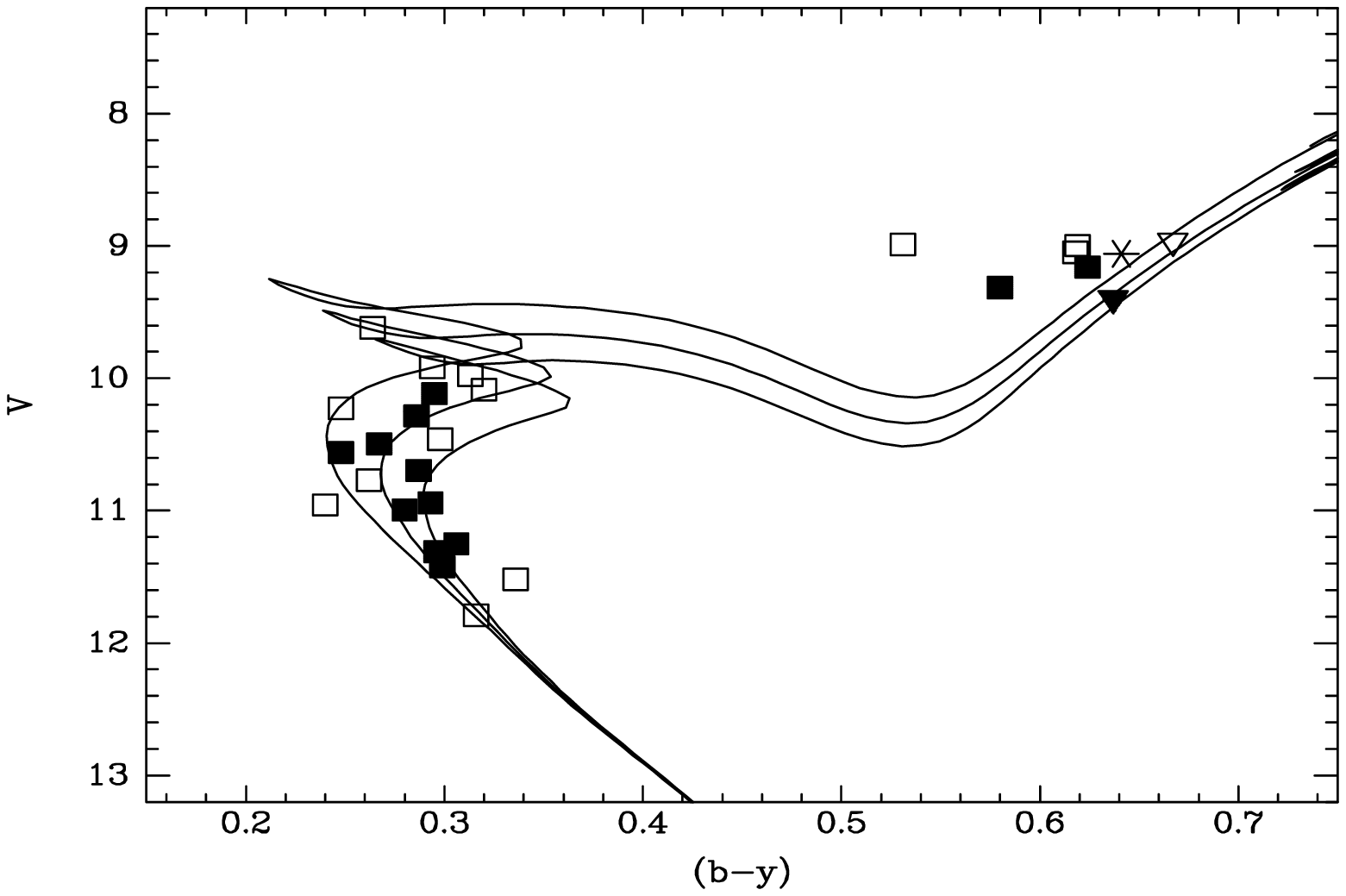]{Str\"omgren $by$ color-magnitude diagram for NGC 752 presenting data from Tables 1 and 2 compared to the same isochrones displayed in Figure 1 except that the $(b-y)_0$ colors of the isochrones have been reddened by 0.027 to match the data.  \label{fg2}}
\clearpage


\begin{deluxetable}{lccccccccccc}
\tablenum{1}
\tablecolumns{12}
\tablewidth{0pc}
\tablecaption{Extended Str\"omgren Photometry for Giants in NGC 752}
\tablehead{
\colhead{I.D.} & \colhead{$V$} & \colhead{$b-y$}   &
\colhead{$m_1$} & \colhead{$c_1$} & \colhead{$hk$} &
\colhead{$\sigma_V$} & \colhead{$\sigma_{by}$} & \colhead{$\sigma_{m1}$} &
\colhead{$\sigma_{c1}$} & \colhead{$\sigma_{hk}$} & \colhead{No. Obs.} }
\startdata
H 110 & 8.988 & 0.531 & 0.266 & 0.467 & 0.815 & 0.009 & 0.000 & 0.002 & 0.012 & 0.029 & 2 \\
H 213 & 9.049 & 0.618 & 0.362 & 0.430 & 1.073 & 0.007 & 0.005 & 0.012 & 0.005 & 0.022 & 4 \\
H 75 & 8.996 & 0.619 & 0.362 & 0.430 & 1.079 & 0.011 & 0.026 & 0.039 & 0.020 & 0.033 & 2 \\
H 27 & 9.159 & 0.624 & 0.368 & 0.418 & 1.091 & 0.009 & 0.009 & 0.013 & 0.015 & 0.012 & 1 \\
H 295 & 9.315 & 0.580 & 0.369 & 0.394 & 1.063 & 0.001 & 0.003 & 0.016 & 0.023 & 0.023 & 2 \\
H 77 & 9.395 & 0.637 & 0.376 & 0.451 & 1.098 & 0.003 & 0.004 & 0.005 & 0.017 & 0.012 & 2 \\
H 208 & 8.962 & 0.667 & 0.407 & 0.384 & 1.120 & 0.013 & 0.005 & 0.012 & 0.020 & 0.006 & 3 \\
\enddata
\end{deluxetable}

\clearpage


\begin{deluxetable}{lcccccccccccc}
\tabletypesize\small
\tablenum{2}
\tablecolumns{9}
\tablewidth{0pc}
\tablecaption{Extended Str\"omgren Photometry for Main Sequence Stars in NGC 752}
\tablehead{
\colhead{I.D.} & \colhead{$V$} & \colhead{$b-y$}   &
\colhead{$hk$} &
\colhead{$\sigma_V$} & \colhead{$\sigma_{by}$} & 
\colhead{$\sigma_{hk}$} & \colhead{Nts Obs.} & \colhead{Note} }
\startdata
H 58 & 10.498 & 0.267 & 0.445 & 0.013 & 0.005 & 0.033 & 2 & FV\\
H 64 & 10.560 & 0.248 & 0.430 & 0.001 & 0.009 & 0.023 & 2 & FV\\
H 66 & 10.939 & 0.293 & 0.488 & 0.012 & 0.002 & 0.023 & 2 & FV\\
H 105 & 10.282 & 0.286 & 0.453 & 0.002 & 0.006 & 0.001 & 2 & FV\\
H 126 & 10.116 & 0.295 & 0.474 & 0.006 & 0.011 & 0.013 & 1 & FV\\
H 135 & 11.250 & 0.306 & 0.456 & 0.010 & 0.003 & 0.011 & 2 & FV\\
H 139 & 11.792 & 0.316 & 0.473 & 0.008 & 0.011 & 0.012 & 2 & \nodata \\
H 187 & 10.459 & 0.298 & 0.441 & 0.008 & 0.013 & 0.024 & 2 & SB \\
H 189 & 11.308 & 0.296 & 0.447 & 0.004 & 0.018 & 0.033 & 2 & FV\\
H 192 & 10.770 & 0.262 & 0.473 & 0.001 & 0.006 & 0.03 & 2 & Fm\\
H 193 & 10.226 & 0.248 & 0.445 & 0.001 & 0.006 & 0.009 & 2 & \nodata \\
H 205 & \phn 9.919 & 0.294 & 0.436 & 0.005 & 0.012 & 0.022 & 2 & SB \\
H 206 & 10.086 & 0.320 & 0.436 & 0.044 & 0.029 & 0.013 & 2 & SB \\
H 209 & \phn 9.766 & 0.033 & 0.286 & 0.012 & 0.004 & 0.016 & 2 & BS \\
H 222 & 10.993 & 0.280 & 0.391 & 0.006 & 0.001 & 0.000 & 2 & FV\\
H 234 & 10.697 & 0.287 & 0.455 & 0.004 & 0.002 & 0.026 & 2 & FV\\
H 235 & 11.518 & 0.336 & 0.506 & 0.011 & 0.012 & 0.014 & 1 & SB, EB \\
H 238 & \phn 9.979 & 0.313 & 0.465 & 0.003 & 0.002 & 0.015 & 2 & SB \\
H 254 & 10.957 & 0.240 & 0.414 & 0.011 & 0.012 & 0.014 & 1 & SB \\
H 259 & 11.422 & 0.299 & 0.393 & 0.009 & 0.012 & 0.014 & 1 & FV\\
H 300 & \phn 9.619 & 0.264 & 0.478 & 0.009 & 0.001 & 0.020 & 2 & SB \\
\enddata

\end{deluxetable}

\enddocument